\input{epsf}

\documentstyle[11pt,paspconf]{article}

\def\hi{{\sc Hi}}

\begin{document}
\title{Interactions between massive dark halos and warped disks}
\author{Konrad Kuijken}
\affil{Kapteyn Institute, PO Box 800, 9700 AV Groningen, The Netherlands}

\begin{abstract}
The normal mode theory for warping of galaxy disks, in which disks are
assumed to be tilted with respect to the equator of a massive,
flattened dark halo, assumes a rigid, fixed halo. However,
consideration of the back-reaction by a misaligned disk on a massive
particle halo shows there to be strong coupling leading to efficient
damping (or in some circumstances excitation) of the misalignment, and
hence the warp. We therefore discuss possible alternative explanations
of the warp phenomenon, with emphasis on the effect of a responsive,
gravitationally live massive galactic halo.
\end{abstract}

\section{Introduction}

The majority of large \hi\ disks studied to date are warped: the plane
of the gas tilts as a function of galactocentric radius (Bosma
1991). Furthermore, there are regularities to this warping: the inner
parts of the disk (typically within $R_{25}$) are coplanar, then
between this radius and the Holmberg radius the disk often warps with
a common line of nodes, and then at yet larger radii the line of
nodes tends to form a leading spiral (Briggs 1990). The \hi\ layer of
the Milky Way is known to be warped as well: while its amplitude is
small (ca.\ 5$^\circ$ maximum tilt, Burton 1988), it follows the same
pattern out to a radius of 1.5$R_0$. At larger radii, however, the
plane is no longer simply tilted, but distorted: the outermost \hi\ in
the Milky Way disk is displaced towards the NGP by about 1kpc. Several
other examples of regular and irregular warps exist.

As with spiral structure, warps present a winding problem: because of
differential precession of bending wave packets at different radii,
large-scale bending waves in a galaxy disk will soon wind up, in
conflict with the observed straight line of nodes for the inner
regions of the disk. In this case, also, self-gravity of the disk has
been invoked as a mechanism for halting the differential
winding. Sparke \& Casertano (1988), building on the work of Hunter \&
Toomre (1969), showed that a disk in the equator of a flattened halo
could support standing bending waves provided the disk was not too
extended, or the halo too flattened. [Isolated cold disks are not able
to support such waves: unless the disk surface density $\Sigma$ is
truncated at the outer edge of the disk such that $\Sigma(R)\propto
(R_{\rm out}-R)^\alpha$ with $\alpha<1$, bending waves will slow down
and dissipate before reaching the edge and being able to reflect off
it.]

The Sparke \& Casertano modes are large-scale, global modes of the
disk, which makes it instructive to step away from the local WKB
approximation and to consider the dynamics of the entire system. This
can be done most easily by considering a galactic disk as consisting
of a collection of concentric spinning rings, able to tilt with
respect to each other, and moving in the combined gravity field of the
other rings and a surrounding massive halo. It can be shown that a set
of stars on initially circular orbits behaves in the same way as such
a ring under the action of torques. Such a system is clearly in
equilibrium when all the rings are co-planar, and moreover, if the
halo is spherical, a new equilibrium is obtained when all rings are
tilted through the same angle. Sparke \& Casertano recognised the
normal modes they calculated as a continuation of this trivial mode to
non-spherical halos.  The halo exerts a net torque on the disk once it
is tilted, causing an overall precession of the spin axis of the disk;
and by allowing the disk to warp, the precession speeds of the rings
can all be adjusted to that same common value. [Rings which would
naturally precess more slowly than the overall pattern need to be
further from the halo equator, so that the torque on them can be
augmented by the gravity of the remaining, less-inclined parts of the
disk.] The precession speed $\Omega_p$ of the pattern is also simply
obtained as the ratio of the torque exerted by the halo (with circular
\& vertical frequencies $\Omega_H(R)$ and $\nu_H(R)$) to the spin of
the disk (given by the total angular velocity of the disk matter
$\Omega(R)$):
\begin{equation}
\label{eq-scomegap}
\Omega_p={\int dR R^3\Sigma(R) (\Omega_H^2-\nu^2_H) \over \int 2dR 
\Omega(R)R^3\Sigma(R)}
\end{equation}
The numerator in this expression is the overall torque from the halo
experienced by the disk if it is inclined by unit angle, and the
denominator the torque exerted by the coriolis force in the frame
rotating with the precession frequency of the mode $\Omega_p$.  The
balance as a function of radius of the coriolis and halo torques
determines the shape of the warped equilibrium. If the coriolis torque
dominates over most of the disk, the warp will be away from the halo
equatorial plane (type~I in SC's notation), otherwise the outer parts
of the disk will warp down towards the halo equator (type~II).

The linear analysis of Sparke \& Casertano can be extended to
non-linear tilt amplitudes (Kuijken 1991): such calculations reveal
that realistic amplitudes can indeed be attained with models of this
kind, though the maximum amplitude of the type~I warps is rather
small.

\section{Back-reaction on the halo}

A possible worry (Toomre 1983; Binney 1992) is that the halo has been
assumed to be a fixed 'potential bath' for the disk to precess in. At
least, this assumption violates conservation of angular momentum of
the system as a whole, unless the halo is considered to have an
infinite moment of inertia. More crucially, if real dark halos are
stellar dynamical entities, they will respond to gravitational fields
on dynamical time scales, and in particular the central regions of a
halo may be expected to respond to the disk precessing through it.

Typical precession speeds for a SC normal mode are
$\Omega_p\sim-\epsilon V_c/R_d$, where $V_c$ and $R_d$ are the
circular speed and half-mass radius of the disk and $\epsilon$ is the
flattening of the halo potential; for typical halo flattenings of 0.1
(much smaller amplitudes will not generate warps of the observed
amplitudes) the precession of the disk might be expected to provoke a
significant response of the dark halo at radii inside $10R_d$.

The importance of this back-reaction may be assessed in two ways:
either by considering the effect of the precessing disk on halo
orbits, from which a first-order halo over-density (and hence the
gravitational effect of this over-density, or wake, on the disk
precession) can be derived; or from numerical simulations, which give
fully self-consistent, but less general, results. The first approach
has been taken by Nelson \& Tremaine (1995), and the second approach
by Dubinski \& Kuijken (1995). Results agree: the effects of halo
response to the disk precession are {\em strong}, operating on the
dynamical time-scale of the halo. As a consequence, halos cannot be
viewed as rigid background sources for a torque, but it is essential
to view the disk-halo system as one, coupled, whole.

\begin{figure}
\centerline{\epsfbox{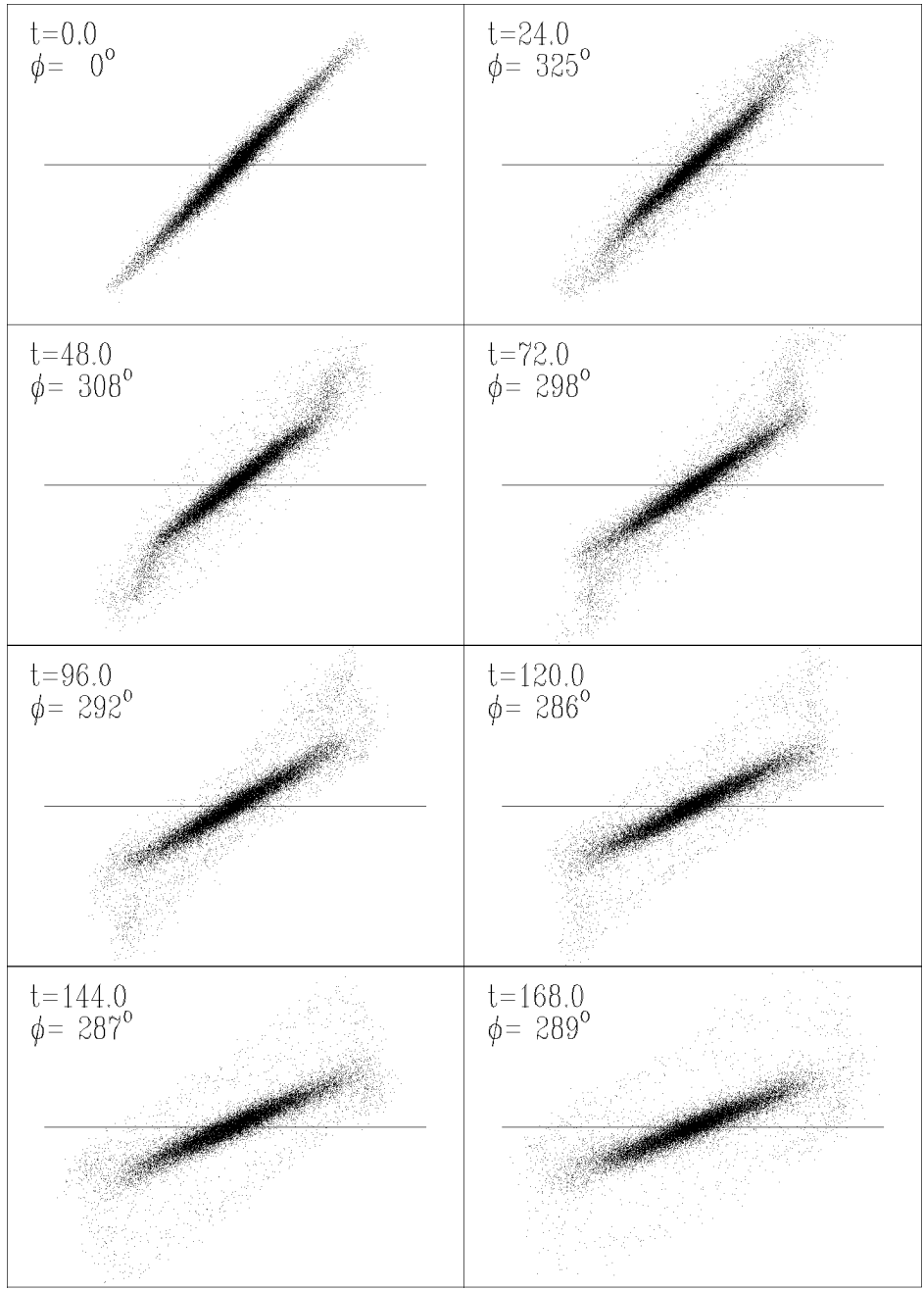}}
\caption{N-body simulation of a stellar-dynamical disk in a live,
non-rotating particle halo (Dubinski \& Kuijken 1995). Only disk
particles are shown. The precession of the disk is rapidly halted as a
result of the halo aligning with the disk, showing that the warp modes
derived assuming rigid background halos in fact are damped on
dynamical time-scales.}
\label{fig-nbody}
\end{figure}

A representative numerical simulation of the evolution of a tilted,
precessing disk in a flattened halo is illistrated in
Figure~\ref{fig-nbody}. Even though initially the precession is set to
agree with the SC formula, the halo responds quickly to the misaligned
disk, and the inclination of the disk with respect to the halo decays
before the disk has been able to complete a single precession. Since
dynamical times are longest in the outer parts of the system, the
disk-halo alignment spreads outwards, completing on ca. 3 local
orbital times.  Responsive dark halos appear unable to support the
warp modes envisaged by SC. Inaccuracies in the N-body simulations are
not the cause of this finding: when the disk particles are replaced by
a solid, spinning disk whose dynamics are followed with the Euler
equations the same result is found, whereas keeping the halo particles
fixed in space and only allowing the disk particles to evolve in the
passive halo yields the long-lived normal mode predicted by SC.

Nelson \& Tremaine (1995) derive slightly longer time-scales from
their analytic work, presumably because the self-gravity of the halo
response is not included in their calculations. Their calculations
also show that radially anisotropic velocity dispersion in the halo (a
plausible consequence of a collapse formation for the halo) lead to
faster damping than isotropic velocity ellipsoids. Interestingly, they
find that under certain circumstances bending can be excited rather
than damped by the halo, on similarly short time-scales: this can
occur when the halo is prolate (i.e. the disk lies perpendicular to
the longest halo axis), or the halo rotates retrograde to the
disk. The prediction for excitation in counter-rotating halos has
since been verified by Dubinski \& Nelson (in preparation) with N-body
simulations. In any case, only under very special circumstances can a
misaligned, precessing disk survive for a large number of dynamical
times in a dark halo.

\section{Alternative explanations}

Given the problems for the normal mode theory, we are still left
lacking an explanation for this common phenomenon of disk
warping. Promising avenues to possible explanations include:

\subsection{Accretion (Ostriker \& Binney 1989; Binney 1992)}
If isolated galaxies are unlikely to be able to sustain warps for very
long, interaction with the environment perhaps offers a solution. In
closed or critical-density universes, galaxies continue to accrete
matter from ever-larger distances as time goes on. Tidal torquing
between different collapsing galaxies will give this infalling
material some non-zero angular momentum, and as this material merges
into the dark halo, the changing angular momentum will probably result
in some realignment of the halo axes. The time-scale for this
realignment is set by cosmology, and is probably close to the Hubble
time.

How will a disk respond to such realignment? If the disk were rigid,
it would certainly change its orientation under the influence of the
changing tidal halo field it was experiencing. In reality, disks are
floppy, with different natural precession frequencies at different
radii, so this overall disk slewing will be accompanied generically by
warping.

\subsection{Equipartition with a lumpy halo (Nelson \& Tremaine 1996)}
The energy associated with a warp mode excited to realistic amplitude
is found to be comparable to the orbital energy of a massive globular
cluster in the halo. If there is sufficient time for equipartition to
be established, and if the halo is dominated by object in this mass
range, then statistical physics dictates that warps will have to be
excited stochastically to observed amplitudes. The main argument
against this possibility are the standard ones against halo objects of
such high mass: they would also strongly heat the stellar
disk. Furthermore, if the microlensing results reported by Freeman at
this conference are to be believed, halo objects have much smaller
masses.

\subsection{Forcing by satellites (Weinberg 1995)}

M.~Weinberg recently showed that the resonant response of the
Milky Way halo to the orbiting Magellanic Clouds raises sufficient
tides in the halo to have a noticeable effect on the stellar disk;
moreover, it does not take excessively high estimates for the mass of
the LMC to obtain the observed amplitude for the Milky Way's
warp. This result highlights the importance of self-consistent
treatment of the halo response to perturbations, and the strength of
coupling of a wake raised at large radius to the inner regions of the
halo. (Weinberg's model for the halo is a scalefree isothermal halo,
but the same gravitational propagation of the response to smaller radii
would be expected in other halo models too). As a ubiquitous
explanation for warps, this work may be criticized on the grounds that
not all warped galaxies have satellites as important as the LMC, and
that while the warp of the Milky Way is rather low-amplitude compared
to other examples, the mass of the LMC is quite possibly lower than
assumed by Weinberg's calculations. Nevertheless, this striking result
is a vivid reminder of the importance of collective and resonant
effects in galaxy interactions.

\subsection{Halo-enhanced tides}

\begin{figure}
\epsfxsize=7truecm
\centerline{\epsfbox{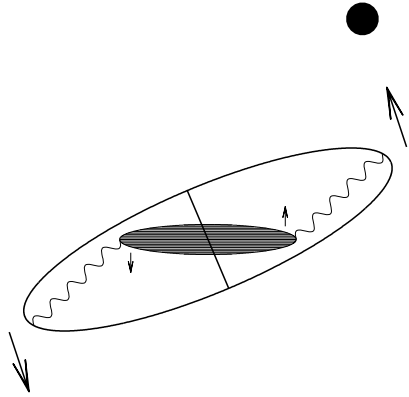}}
\caption{A simple two-component galaxy, consisting of a
spinning, rigid central disk, and a flattened halo, here represented by a
massive outer ring.}
\label{fig-sketch}
\end{figure}
 If not all warped galaxies have companions which may have distorted
them, perhaps infrequent passages by other galaxies may raise
sufficient tides in the halos to warp the disks noticeably. A very
naive model (see Fig.~\ref{fig-sketch}) shows that such an effect
might be important: consider a galaxy as made up of a massive,
spinning inner disk, and a surrounding dark halo which we here model
as a rigid, possibly spinning, concentric ring surrounding it. If we
now tidally perturb such a galaxy with a distant passage, what will
happen to the system? Assuming all tilt angles are small, it is
possible to show that the dynamics of symmetry axes of the disk and
surrounding `halo' are governed by the equations
\begin{eqnarray}
\ddot x_d&=&-S_d y_d +C m_h(x_h-x_d)\\
\ddot y_d&=&\phantom{-} S_d x_d +C m_h(y_h-x_d) + T(t) I_d\\
\ddot x_h&=&-S_h y_h +C m_d(x_d-x_h)\\
\ddot y_h&=&\phantom{-} S_h x_h +C m_d(y_d-x_h) + T(t) I_h.
\label{eq-tilts}
\end{eqnarray}
Here $(x_d,y_d)$ and $(x_h,y_h)$ are the orientations of the
polar axis of the disk and halo, where $x_d=\theta_d\cos\phi_d$, etc.\
for spherical polar coordinates $(\theta_d,\phi_d)$.

The coefficient $C$ expresses the gravitational coupling between the
halo and the disk, $T(t)$ is the strength of the tidal torque
(the perturber is assumed to move uniformly along the $z$-axis), and
$I_d$ and $I_h$ are the moments of inertia of disk and `halo'.

\begin{figure}
\epsfxsize=\hsize
\epsfbox{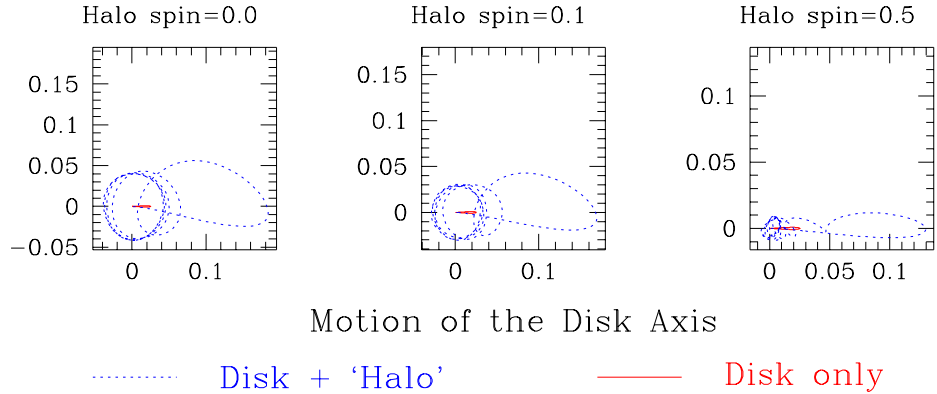}
\caption{The response of a two-component galaxy, consisting of an
inner massive disk and an outer heavy ring to the tide from the
passage of a perturber. The motion of the spin axis of the disk is
plotted with and without the gravitational coupling $C$ between disk
and halo taken into account, for `halos' of different spins. Without
coupling, the disk axis executes a small wobble, but once the
gravitational disk-halo coupling is included the response of the outer
halo significantly enhances the tide experienced by the disk. Scales
are arbitrary.}
\label{fig-response}
\end{figure}

In the absence of the halo ($C=0$), the disk will respond to the
tidal field, and execute a small precessing motion.  After the passage
is over, overall energy and angular momentum conservation of the disk
return it in its original state. [A real disk would warp during the
reaction to the tidal field, but eventually the bending waves would
dissipate and the disk would return to its original orientation.]
However, if we include the coupling between the disk and the halo, the
result becomes very different.  The halo, having a larger radius and a
smaller spin than the disk, responds more strongly to the tidal field
than the disk. As it tilts, however, it exerts its own tidal field on
the disk, and, because of the halo's proximity to the disk, this field
is stronger than the external one. The net result is that the halo
mediates the external field, and transmits it onto the disk more
strongly than the original one. Depending on the rotation of the halo,
this response is stronger or weaker: simple numerical experiments with
equations (\ref{eq-tilts}) show that a factor of five enhancement is
not difficult to attain (Fig.~\ref{fig-response}).

Less drastically simplified models of the halo give comparable
results (Lynden-Bell 1985; see also Nelson \& Tremaine 1996).

\section{Summary and Conclusions}

The phenomenon of warps, affecting many disk galaxies, continues to
puzzle. The promising Sparke-Casertano model that warps reflect a
long-lived misalignment between a disk and a flattened massive halo
appears to fail once the response of the halo to the precession of the
disk is taken into account (Nelson \& Tremaine 1995; Dubinski \&
Kuijken 1995), leaving us to search for other possibilities. Perhaps
the answer lies in magnetic generation of warps (Battaner et al.,
1990; see Binney 1991 for a critical discussion of this possibility),
but other gravitational possibilities remain. In this article, I have
tried to stress the point that the dark halo should be considered an
integral, dynamical part of the galaxy, rather than just a potential
energy bath. With the discovery by Weinberg (1994) that spherical
models can undergo very weakly damped modes, goes the implication that
once disturbed, halos can ring for a large number of dynamical times,
forcing oscillations and warps in embedded galaxy disks. Combined with
the enhanced excitation of tides on the disk by the halo, perhaps most
galaxy warps are such weakly-damped, tidally induced ringing?

How can these ideas be tested further?

We still lack a clear observational study of the statistics of
warps. In particular, it would be good to know how regular warps are,
if there is any relation between warps and bars, warps and
lopsidedness, kinematic peculiarities and warps, links with
environment, etc. On the theoretical side, simulations of some of the
suggested scenarios would be very useful: in particular, the details
of the response of the halo, and hence the disk, to secondary infall
are difficult to predict and would benefit from such a study.

It is clearly not yet time for warps to stop challenging people's
minds!

 \end{document}